\documentclass[journal]{IEEEtran}
\usepackage{graphicx} 
\usepackage{hyperref}

\usepackage{multirow}
\title{Space-Air-Ground Integrated Network (SAGIN): A Survey}

\author{Jiming Chen\\~\IEEEmembership{School of Integrated Circuit Science and Engineering 
(Exemplary School of Microelectronics)\\ University of Electronic Science and Technology of China}\\ \hspace*{\fill} \\
Han Zhang\\~\IEEEmembership{School of Integrated Circuit Science and Engineering 
(Exemplary School of Microelectronics)\\ University of Electronic Science and Technology of China}\\\hspace*{\fill}\\
Zhe Xie\\~\IEEEmembership{School of Electronic Science 
and Engineering\\University of Electronic Science and Technology of China}
}

\date{July 2023}

\begin{document}

\maketitle

\begin{abstract}
Since existing mobile communication networks may not be able to meet the low latency and high efficiency requirements of emerging technologies and applications, novel network architectures need to be investigated to support these new requirements. As a new network architecture that integrates satellite systems, air networks and ground communication, Space-Air-Ground Integrated Network (SAGIN) has attracted extensive attention in recent years. This paper summarizes the recent research work on SAGIN from several aspects, with the basic information of SAGIN first introduced, followed by the physical characteristics. Then the drive and prospects of the current SAGIN architecture in supporting new requirements are deeply analyzed. On this basis, the requirements and challenges are analyzed. Finally, it summarizes the existing solutions and prospects the future research directions.
\end{abstract}

\begin{IEEEkeywords}
Space-Air-Ground Integrated Network, Non-Terrestrial Network, Network Design and Protocol Optimization, Resource Allocation, Performance Analysis.
\end{IEEEkeywords}

\section{Introduction}
Over the past few years, the rate of technological and scientific progression has been unprecedented, yielding conspicuous impacts on the diverse facets of communications, ranging from User Experience (UE) to the service and applications to be supported. As the 5th generation mobile communication network (5G) is fully popularized and commercialized, researchers are focusing on developing better communication technologies for the purpose of building a cross-region, cross-air-space and cross-sea integrated network systems to achieve a truly seamless global coverage network. Whereas, current communication systems, such as 5G Non-Terrestrial Network (NTN) cannot support the aforementioned hypothesis, mainly due to its poor access flexibility and high latency \cite{sheng2021space}.

Therefore, a brand new network architecture called Space-Air-Ground Integrated Network (SAGIN) is proposed. Taking advantage of the evolving technology of satellite, carriage, Unmanned Aerial Vehicle (UAV), artificial intelligence (AI), security and computing power networks, SAGIN integrates satellites, airborne platforms (airship, balloon, UAV, etc.), and terrestrial networks. Similar to NTN technology, which absorbs the dual advantages of traditional satellite communication and ground mobile communication, SAGIN utilizes its heterogeneity to optimize its performance in global coverage, high reliability and throughput \cite{cui2022space}. Endowed with great potential and value, SAGIN will bring revolutionary changes and impacts to human society. The integration of these network segments will bring many benefits to future 5G and 6G communication and service development.

SAGIN can be used in a multitude of practical fields, compromising earth observation and mapping, Intelligent Transportation System (ITS), military mission, disaster rescue and so on \cite{liu2018space}. Besides, with SAGIN architecture, 6G is anticipated to achieve the better UE in all the scenario with the requirements with wide-area broadband access, large-scale connectivity, time-sensitive connectivity, high-precision positioning such as emergency communications, IoT in wide area, ecological remote sensing monitoring and Industrial Internet of Things (IIoT) (connect factories and machines). The analysis and optimization of SAGIN will be presented later in the passage.

The reminder of this paper is organized as follows. Section \ref{background} introduces three main portions of SAGIN architecture and compares with the previous networks structures. Section \ref{physical} present the physical layer characteristics and possible problems at the technical level. We analyze diversified applications and corresponding optimization methods of SAGIN in Section \ref{Requirements} to Section \ref{Challenges}. Finally, we summarize the whole paper in the last section. 

\section{Background}
\label{background}
In this section, we first give an overview of the integrated network system architecture and then we provide a brief comparison of previous techniques.
\subsection{System Architecture}
SAGIN mainly contains three segments: space, air, and ground. Integrated through heterogeneous network, these segments can work independently or inter-operationally \cite{liu2018space}, in order to achieve global coverage and seamless connection, providing more universal and balanced communication services for any place, any time and any device.
\subsubsection{Space Network}
A space network consists of satellites in different orbits and their corresponding ground infrastructure such as ground base stations, network operators and control centers, which have different characteristics due to position. Three main components are included in a typical satellite communication system: ground, space segment, and air-ground link. The ground part generally includes various information stations, satellite measurement and control centers, and the corresponding satellite measurement and control networks and network control centers.

According to different heights, satellites can be divided into three categories: geostationary (GEO) satellites, Medium Earth Orbit (MEO) satellites and Low Earth Orbit (LEO) satellites. Together, these satellites form Multilayered Satellite Networks (MLSNs), which is a practical architecture for next generation \cite{wang2022evaluating}. MLSNs are composed of multiple satellite networks and has a hierarchical structure. MLSNs also have multiple link types, such as Inter-Satellite Links (ISL) and Inter-Layer Links (ILL). However, the different height of the satellite determines the coverage capability. The coverage area of a single satellite in high orbit is fixed relative to the ground, and a single satellite can cover 42\% of the Earth's area. Generally, 3-4 satellites can complete the global coverage of polar regions.

The single satellite coverage area of MEO, which with an orbit height of 2000 km-20000 km and covering 12\%-38\% of the earth's surface, is much smaller than that of GEO. Therefore, it takes a dozen or more satellites to form a star chain to complete global coverage. MEO are designed to provide high-bandwidth, low-cost, low-latency satellite Internet access, which with transmission latency of less than 150ms and system capacity up to 15Gbps. Similarly, LEO is of lower cost and smaller coverage, requiring multiple satellites to form star chain to achieve global coverage.

\subsubsection{Air Network}
Air network is air mobile systems that uses aviation equipment (mainly aircrafts, balloons and UAVs) as the carriers to collect, transmit and process information. Similarly, air network can be categorized into High Altitude Platforms (HAPs) and Low Altitude Platforms (LAPs). Compared with terrestrial network base stations, air networks are easy to deploy, have wide coverage, and can provide regional wireless access. At the same time, the cost is much lower than that of space networks only.
Aircrafts and balloons are the main infrastructure for HAPS, providing broadband wireless communications to complement terrestrial networks. In recent years, many research have actively considered HAPS as a viable technology for future wireless communication networks, to provide wireless services directly to terrestrial network users due to its low latency.

HAPS systems have played different roles in the models proposed in several studies. These roles can be divided into two main parts: the connection with the ground and the connection with the satellite. When connected to the ground, HAPs layers can provide urban, suburban, and remote areas with fast Internet access and wireless communication services such as Internet of Things (IoT) and distributed machine learning, thereby reducing dependence on terrestrial and satellite networks. When connected to the satellite, it can act as a distributed data center to record the orbit path of the satellite, monitor the joint warning, and calculate the possibility of collision between satellites. Providing this information to different satellite companies in time is crucial to maintain the well function of the satellite starlink.

The typical infrastructure of LAPs is UAV, which has more flexibility in deployment, application and maintenance, than that of HAPs. According to proper scheduling strategy, LAPs can provide lower resource overhead in communication.

It is worth noting that UAV networks have great potential in air networks. UAVs have been widely used in both military and commercial applications. The complex multi-UAV network can be applied into the SAGIN system, which has great advantages in survivability and reliability. With the cooperative communication, shared information and distributed processing ability of UAVs, the multi-UAV network further reduces the system delay and improves the flexibility. The multi-UAV network can be used as an air base station or mobile cloud computing system.

\subsubsection{Ground Network}
Ground networks are terrestrial communication systems, including cellular network, mobile and hoc network (MANET)(a type of self-organizing network without a center and no physical base stations are required, hosts are connected to each other, and hosts can act as servers), and wireless local area networks (WLANs).

With the maturity of 5G, base stations of cellular mobile networks on the ground may be directly incorporated into the SAGIN system. But now more and more scholars believe that with the deployment of air and space networks, the whole structure can be further optimized. In the existing research, 5G NTN architecture based on Software Defined Network (SDN) and network function virtualization (NFV)\cite{barakabitze20205g} is proposed. At the same time, concurrent multipath transmission techniques and analytical models are proposed to achieve the optimal separation of multipath applications between space and segment \cite{wang2022conditional}. SAGIN includes a unified terminal and access protocols, building an integrated control and data plane via Radio Access Network (RAN) and Core Network (CN).  

\subsubsection{Comparisons among these segments}
Each segment has its own strength and limitations. The table\ref{TableSystemComparision} below demonstrates delays, advantages and disadvantages of different systems.

\begin{table*}[]
\centering
\caption{Table of Systems Comparison}
\label{TableSystemComparision}
\renewcommand{\arraystretch}{1.3}
\begin{tabular}{|c|c|c|c|c|}
\hline
Segment & Objects & One-way delay  & Advantages & Disadvantages \\ \hline \hline
\multirow{3}{*}{Space} &
  GEO &
  About 270ms &
  \multirow{3}{*}{\begin{tabular}[c]{@{}c@{}}Large coverage   \\ Broadcast/multicast\end{tabular}} &
  \multirow{3}{*}{\begin{tabular}[c]{@{}c@{}}Long propagation latency   \\ Limited capacity   \\ High mobility\end{tabular}} \\ \cline{2-3}
        & MEO     & About 110ms    &            &               \\ \cline{2-3}
        & LEO     & Less than 40ms &            &               \\ \hline
\multirow{3}{*}{Air} &
  \multirow{3}{*}{\begin{tabular}[c]{@{}c@{}}Airship\\   Ballon\\    UAV\end{tabular}} &
  \multirow{3}{*}{Medium} &
  \multirow{3}{*}{\begin{tabular}[c]{@{}c@{}}Wide coverage    \\ Low cost   \\ Flexible deployment\end{tabular}} &
  \multirow{3}{*}{\begin{tabular}[c]{@{}c@{}}Less capacity\\    Unstable link\\ High mobility\end{tabular}} \\
        &         &                &            &               \\
        &         &                &            &               \\ \hline
\multirow{4}{*}{Ground} &
  \multirow{4}{*}{\begin{tabular}[c]{@{}c@{}}Cellular\\    Ad Hoc   \\ WiMAX\\    WLAN\end{tabular}} &
  \multirow{4}{*}{Lowest} &
  \multirow{4}{*}{\begin{tabular}[c]{@{}c@{}}Rich resources\\ High throughput\end{tabular}} &
  \multirow{4}{*}{\begin{tabular}[c]{@{}c@{}}Limited coverage\\    Vulnerable to disaster\end{tabular}} \\
        &         &                &            &               \\
        &         &                &            &               \\
        &         &                &            &               \\ \hline
\end{tabular}
\end{table*}

\subsection{Comparison with Previous Techniques}
SAGIN is considered to be the key technology in the future 6G era, which has the advantages of lower latency and higher flexibility compared with the previous communication models. This section analyzes the characteristics of SAGIN by specifically comparing the characteristics of each generation of communication network architecture.

\subsubsection{4G and Before}
The first generation of mobile phone systems utilized cellular network technology. The concept of cellular was proposed by Bell LABS and studied in many parts of the world in the 1970s. The first cellular system in the United States, Advanced Mobile Phone Service (AMPS)\cite{mac1979advanced}, became a reality in 1979.

When the mobile communication technology comes to the digital communication era, the frequency of data transmission has become the most fundamental factor to improve efficiency. There was a continuous progress from 1G to 4G technologies in terms of network architecture, mainly focused on cellular system\cite{suleiman2022investigating, lou2021green}.

At the same time, satellite communication technology is constantly evolving. As the number of satellites launched surged, space-ground integrated systems became more diverse and powerful \cite{wang2022ultra}.  Among those systems, there are mainly two kinds of network architectures, one is GEO system such as the Transformational Communications Satellite (TSAT) system\cite{pulliam2008tsat}, and the other is non-GEO (NGEO) system such as O3b\cite{wood2014revisiting}, Iridium\cite{maine1995overview}, Globalstar\cite{dietrich1998globalstar}, etc.

Although there have been many studies to integrate terrestrial networks with satellite networks, due to various factors, NTN were not covered in 4G and previous protocols\cite{borgaonkar2018new}. In most of the scenarios, the two systems are more relatively independent and play their respective roles.

\subsubsection{5G NTN}
In the 5G era, the terrestrial networks were further developed based on the cellular system architecture of the 2G and 3G, which splits the base station into a Centralized Unit (CU) and a Distributed Unit (DU)\cite{lin20215g,zhao2022wknn}. Through three different levels, CN, CU and DU, resource scheduling and load balancing were better realized.

In addition, NTN was integrated into 5G communication protocols by 3GPP R17 version\cite{lin20215g}, which is one of the advantages that distinguish 5G from 4G and before.

NTN is an important supplement to terrestrial cellular systems.  Utilizing the integration of satellite network and terrestrial 5G network, it can provide ubiquitous coverage with few limitation by landform, connect with space, ocean and ground, to form an integrated ubiquitous access network, enabling full scene on-demand access.

\subsubsection{6G and Future}
Under the background of the era of AI, there are complex application scenarios and higher requirements for information transmission than in the past. 6G network architecture has attracted great attention from the academia \cite{letaief2019roadmap}. SAGIN architecture becomes one of the most promising technologies to realize the 6G. By integrating the heterogeneous network between the three network segments, it is easy to build a hierarchical broadband wireless network, so as to make good use of the advantages of those three network segments than separately.

\section{physical layer character}
\label{physical}
In this section, we explore the technical difficulties in the development of SAGIN technology through the analysis of the physical layer of some SAGIN structures.

\subsection{frequency bands}
In the selection of frequency bands, we need to pay attention to two factors: one is the size of this layer of communication information, and the other is to consider the impact of different frequencies on the loss rate of electromagnetic waves. From the perspective of digital communication principles, the higher frequency of the signal, the higher efficiency of its theoretical information transmission. The difference between SAGIN technology and traditional communication is that its air-ground communication is extremely frequent, resulting in signal interference and spectrum allocation problems that need to be solved urgently.

In fact, the International Telecommunication Union (ITU) has regulated many existing frequency bands. In general, inter-satellite communications have higher frequency bands, while terrestrial communications are generally narrower due to environmental constraints. Having considered various factors such as the environment, signal interference and frequency band utilization efficiency, scholars have sorted out the following table through researches\cite{liu2018space}.

\begin{table*}
\centering
\caption{Table of Frequency Bands}
\label{TableFrequencyBands}
\renewcommand\arraystretch{1.3}
\begin{tabular}{|c|c|c|}\hline
Frequency bands & Frequency range & Lentended service                                         \\ \hline \hline
L               & 0.39-1.55       & GPS, satellite phone,space-air and air-ground communication \\ \hline
Ku              & 12.5-16         & broadcast satellite service                                \\ \hline
Ka              & 26.5-36         & close-range targeting radars on military aircraft          \\ \hline
Q               & 36-46           & High throughout satellite service                          \\ \hline
V               & 46-56           & High throughout satellite service                          \\ \hline
W               & 75-100          & High throughout satellite service     \\        
\hline
\end{tabular}
\end{table*}

From the table\ref{TableFrequencyBands}, it is not difficult to conclude that the L band can be applied not only to satellite-ground communication, but also to space-air communication, as well as air-ground communication. 

Ku-, Ka-, Q/V-, and other socalled millimeter wave (mmWave) bands have the highest propagation efficiency, which satellite communications are utilizing. It should be noted that higher wavelengths cannot be adapted to air-ground communication because their loss rate is too high. According to the propagation and attenuation characteristics of signals of different frequency, ITU divides frequency bands according to the altitude, which improves the efficiency of the SAGIN system.

\subsection{Propagation Channel}
The SAGIN system primarily relies on wireless communication, which means there are high variance in time, frequency, and space When signals penetrate the propagation medium. What's more, high mobility and long propagation distance of the SAGIN system itself make it distinctive. For instance, the high Bit Error Rate (BER), bandwidth asymmetry, intermittent interruption and other factors in satellite channel will affect the UE \cite{wang2023resident}. Due to the long distance between satellite transponders and ground terminals in satellite communication systems, satellite communications are seriously affected by multipath fading and shadow effects. These factors seriously affect the quality of communication and the spectral efficiency of the system. 

In view of this problem, some scholars have proposed the transmission characteristics of satellite communication channels and analyzed the simplified fading channel model\cite{liu2016analysis,wang2022stochastic}. The model is based on C. Loo models and Corazza models to describe received signals, multipath fading, and shadow effects. The simulation results show that the model dynamically switches between ideal channel state and imperfect channel state. This study is of great significance for the implementation of actual channel simulation and can be widely used in satellite communication system research.

SAGIN, which is reflected in the communication between satellites and the ground or between air and the ground, is also quite different from previous satellite phones. SAGIN consists of many communication nodes, and these nodes are dynamic (only GEO is fixed relative to the ground).The physical layer channels of the system change with the changes of the communication nodes in the system, and higher-layer decisions of flow control and routing are also affected, which in turn affects the latency and data throughput of the entire system.There are many improvement measures in academia, and these methods have been improved from the bandwidth of the physical layer, the packet switching algorithm of the data link layer, and the optimization of the routing path at the network layer.

The principle of cross-layer optimization spanning from physical layer to the network layer should be considered for SAGIN design. Specific optimization measures and technical difficulties are mentioned below.

\section{Requirements}
\label{Requirements}
Through the deep exploration of the SAGIN system, it can be found that the drive of the new generation of communication technology such as 6G is also applicable to the SAGIN system, namely application and technology\cite{wang2023reliability}. The following will explain the development requirements of SAGIN from the perspectives of application and technology.

\subsection{Application}
There are four main types of application scenarios in SAGIN: wide-area broadband access, wide-area large-scale connection, wide-area time-sensitive connection, and wide-area high-precision positioning.

\subsubsection{Wide-area Broadband Access}
Wide-area broadband access refers to providing broadband access to people, especially in remote areas. airplanes, drones, cars, etc., are utilized to narrow the digital divide, reduce energy consumption, and to a certain extent, promote the coordinated development of other fields such as economy and education. At present, the land coverage of mobile communication in the world is still limited. Considering the cost of equipment construction, maintenance and the expected economic benefits, so as to the ability to cover areas such as deserts, oceans, remote mountain areas and polar regions, the communication development of relatively underdeveloped rural areas lags behind. The SGAIN can achieve global three-dimensional coverage and wide-area broadband access at any time and anywhere. It can expand the communication coverage and solve the difficulty of low communication quality in specific areas under the premise of uneven regional investment of operators in ground base stations and other facilities.

\subsubsection{Wide-area Large-scale Connectivity}
Wide-area large-scale connectivity refers to providing connectivity for scenarios such as crop monitoring, depopulated zones monitoring, offshore buoy information collection, ocean-going container information collection, and no-man's land exploration/detection. In the future, intelligent systems such as autonomous vehicles, unmanned logistics systems, and remote robots will be widely deployed in the world and become an important part of human production and life, which has become the current development trend. At that time, information interaction and collaborative work between these intelligent systems will be a ubiquitous scenario. However, due to the limited coverage ability, it is difficult for terrestrial communication systems to provide effective network services for the collaborative work of wide-area intelligent systems. SAGIN can effectively solve such problems.

\subsubsection{Wide-area Time-sensitive Connectivity}
Wide-area time-sensitive connectivity refers to providing network connectivity for delay-sensitive scenarios such as remote smart machine operations, especially for the transportation domain, such as ITS\cite{huang2023system}. Modern transportation needs low energy consumption, high transportation efficiency, comfort, safety and other developments. Realtime intelligent traffic controling and realtime vehicle networking are the future development trends. Traditional mobile communication systems are not enough to support communication services with large capacity, low latency, high reliability and high security. SAGIN reduces the end-to-end delay of remote driving and the return delay of ITS and wire-less measurement system through cooperative satellite and air network.

\subsubsection{Wide-area High-precision Positioning}
Wide-area high-precision positioning refers to providing precise navigation for remote intelligent transportation and high-precision positioning for remote operations. At present, the navigation system centered on the space network and the communication system centered on the ground cellular network are relatively separate. In SAGIN, the two systems can be further coupled. The ultra-high density network of the ground base station is used as the extension and supplement of the ground-based enhanced network of the navigation satellite. At the same time, the rough position and other information of the terminal is obtained by the ground communication network. The combination of terrestrial cellular network and navigation system in SAGIN provides a new opportunity for system-level conductance enhancement and empowerment.

\subsection{Technology}
Technology is another key drive for the development of next generation mobile communication networks. In the past decades, with the development of science and technology, new requirements will be continually generated.

\subsubsection{Satellite Technology}
The development of satellite technology is to be able to adapt to the dynamic changing mission requirements in the future. Satellite manufacturing enterprises have greatly improved their satellite manufacturing capacity and reduced the low-cost development of satellites via various ways such as technology improvement, process optimization and innovation.

LEO constellation navigationt technology is expected to become a new increment of the next generation navigation system\cite{xiao2022leo}. The low orbit altitude of LEO satellite, short signal transmission path and high landing power can improve the positioning effect under occlusion conditions and anti-interference ability. At the same time, the geometric configuration of LEO satellite changes rapidly, which reduces the positioning convergence time and improves the user experience.

Many startups and scalable growth companies are investing in LEO satellite technology, deploying LEO small satellite constellations to enable faster Internet connections. Companies that are growing are meeting the ever-increasing demand for connectivity speeds with super-capacity satellites in GEO and LEO orbits. Satellite technology improves cloud computing, IoT, and backhaul channels of existing terrestrial systems.

\subsubsection{Carriage Technology}
The re-use technology of rockets and flexible satellite launch technology have greatly reduced the cost of space vehicles \cite{jin2018addressing}. Under the premise of making full use of the carrying capacity of rockets, the repeated use of rockets with high-density stacked design can reduce the waste of resources.It greatly improves the satellite launch density, achieves a rocket with multiple satellites, reduces the launch cost, and provides strong support for the construction of large-scale satellite networks in the future. At present, SpaceX can launch 60 satellites per arrow through multiple Falcon 9 rockets.

As more and more LEO satellites and other types of satellites are put into orbit, flexible, on-demand launches are expected. Flexible satellite launch technologies, such as airborne launches using spacecraft, balloons, autonomous launch equipments and unmanned aerial vehicles, have become landable solutions.

\subsubsection{Unmanned Aerial Vehicle Technology}
The advantages of low cost and high flexibility of Uavs will be used in a wider range of scenarios.

With the new battery, hybrid power, ground power supply, wireless charging and other technologies are utilized into the research and development process of UAV \cite{depcik2020comparison}. The endurance ability of UAV is expected to achieve a breakthrough and embark on a new revolution, breaking the current major obstacles restricting the development of UAV.

The application of GPS carrier phase, multiple information sources, UWB wireless positioning and other positioning methods plays a key role in solving major problems such as the safety of UAV flight \cite{lazzari2017numerical}. Combined with advanced speed measurement technology, obstacle avoidance technology, tracking technology, the rapid development of commercial and civilian UAV and other fields, the UAV has gradually acquired the ability to be operated in complex environments.

For the application scenarios that need to transmit massive data, encryption algorithms, authentication mechanisms and other technologies are used to ensure data security, which may cause threats to flight safety and sensitive data leakage.

\subsubsection{AI Technology}
AI technology has been a popular topic in recent years. The fast development of deep learning algorithms, ncrease of massive data access channels, and significant improvements of hardware computing and storage capabilities have led to the explosive development of AI technology \cite{gupta2021fusion}. Breakthroughs have been made in many fields such as speech recognition, image recognition, and emotional communication.

AI technology has a wide range of applications. Not only basic industries such as medical care, agriculture, and urban management are related to AI, but also emerging businesses such as emergency telecommunications and IIoT are gradually inseparable from AI. Today, ChatGPT, a large language model based on deep learning technology, makes it convenient for daily life. Through a large number of natural language processing data sets, neural networks are trained to understand and generate natural language, automatically answer users' questions, provide information and suggestions, and other functions.

AI technology has also achieved initial applications in the communication field, such as reducing network energy consumption and optimizing network deployment through accurate prediction of commercial demands. In the future, AI and communication will be more deeply integrated to solve the problem of complex heterogeneous communication systems \cite{xu2021survey}.

\subsubsection{Security Technology}
SAGIN has the characteristics of open integration, heterogeneous coexistence and ubiquitous connection. Data security is an important bottleneck restricting its development. Blockchain is a promising security technology, which has been rapidly developed in finance, medical care, logistics supply chain and other fields in recent years. Blockchain is composed of asymmetric encryption, distributed computing, peer-to-peer (P2P) network\cite{daniel2022ipfs} and so on. It is decentralized, tamper-resistant and traceable.

Abductive, open and transparent, these characteristics enable blockchain technology to effectively solve the security problems of data storage and transmission.

\subsubsection{Computing Network}
\label{Computing}
Edge computing, which provides computing, storage and network bandwidth close to the data input or user, has been widely used \cite{al2021survey,huang2022wearable}. Its low latency characteristics are not available in traditional solutions. Edge computing can provide more service capabilities and has a wider range of application scenarios. In order to optimize the resource utilization, edge-to-edge collaboration among edge computing, cloud computing and network should be realized, and the computing power allocation and scheduling requirements of edge computing should be optimized.

Data, algorithms, and computing networks have become the new economic elements in the era of AI. By optimizing the allocation of computing resources, the computing power network can carry and process massive data with high efficiency and low latency \cite{huang2021dynamic}. Connecting nodes of computing resources with the network to form an integrated computing service platform is becoming the basis of the development of digital information.

\section{Development Vision}
\label{Development}
SAGIN is typical architecture. Its vision is to meet the growing demand for wide-area intelligent connectivity and global ubiquitous seamless access, establish intelligent connectivity for wide-area objects, provide intelligent services, and provide global uninterrupted and consistent services for human beings. 

It's widely acknowledged that the 5G network coverage is achieved by adding a large number of base stations \cite{lou2023coverage}. The deployment of base stations is limited by physical location, and it is not possible to truly achieve global three-dimensional seamless coverage, in places like oceans, mountains, deserts and other extreme environment and uninhabited areas. Therefore, driven by requirements and technology developments, satellite communications and air communications will be integrated with ground communications. Integrate with different levels ont only terrestrial networks take charge of normalized coverage of urban, but also satellite and airborne realize seamless coverage in remote areas \cite{chen2023coverage}, like dessert and sea. 

\subsection{Typical Features}
SAGIN has three typical characteristics: unified radio interface technology, unified network architecture and unified intelligent control. 

Unified radio interface technology means that satellite communication, air communication and ground communication adopt the radio interface transmission technology under the same framework, the terminal to achieve minimalist and intelligent access. 

Unified network architecture refers to the integrated design of satellite communication, air communication and ground communication under a unified logical architecture and implementation architecture. The network functions can be flexibly divided and intelligently reconstructed to adapt to the characteristics of limited satellite payload resources and dynamic changes in requirements. Although many advantages have been brought in, it also results in that SAGIN is easily affected by limitations of three segments themselves at the same time, such as traffic distribution, spectrum allocation, load balancing, mobility management, power control, route scheduling, end-to-end (E2E) Quality of Service (QoS) requirement\cite{tang2021survey}, etc

Unified intelligent management and control refers to the unified scheduling and control of system resources to achieve global network and green intensification.

\subsection{Core Elements}
The core elements of SAGIN are intelligence and virtualization. In the future, the ubiquitous sensing and computing capabilities from terminals to networks provide a wide range of data and AI. AI will penetrate into all the levels of physical layer algorithms, radio resource management, network function, and service enhancement to realize system intelligence. On the other hand, through the virtualization of space-time frequency radio resources, computing resources, storage resources, interface resources and network functions, the unified resource scheduling and network control of the SAGIN can be realized.

\section{Challenges}
\label{Challenges}
SAGIN system has the characteristics of heterogeneous network, highly dynamic space nodes, time-varying topology, huge spatio-temporal scale, limited resources of space nodes, and attacks on satellite broadcast transmission chain. These characteristics put forward higher requirements for the design of network architecture, integrated communication standards and ISL networking protocol.

\subsection{Dynamic Nodes}
Each node in the traditional terrestrial network is relatively fixed. There is high-speed relative movement between various types of satellites and the ground-based network in SAGIN. If the access and networking of nodes such as HAPs, aircraft, and low-altitude UAVs are considered, the movement characteristics are usually more irregular and the impact will be uncontrollable. One of the main effects caused by the highly dynamic characteristics of nodes is the serious Doppler frequency offset in communication, which leads to high outage rate and high BER of communication links \cite{hoydis2021toward}. These transmission characteristics put forward higher requirements for the design of SAGIN standards.

\subsection{High-speed Topology changing}
In SAGIN, the network is composed of different levels of nodes such as satellites, HAPs, LAPs and ground equipment, which has a three-dimensional architecture completely different from the traditional terrestrial cellular communication network. The movement of nodes in the network will also lead to the high dynamic change of the network topology. On the one hand, the high-speed topology change of the network will cause link changes. It is difficult to transmit data through fixed traffic. On the other hand, network protocols need to consider the contradiction between asymmetric links, quality variation of links and high reliable transmission feedback control under different situations such as multi-hop and relay\cite{liu2022solids}, which will cause low transmission efficiency at the application layer, and even fail to guarantee data transmission quality.

\subsection{Huge Time and Space Span}
Due to the long distance and high frequency, the links between the network nodes are far more than the ground network in time and space span. On the one hand, the transmitted signal between nodes has large attenuation loss and is easily affected by a series of factors such as orbit change, elevation Angle, solar flicker, atmospheric scattering, rain attenuation, and occlusion, resulting in weak received signal and a variety of interference \cite{rummler1981more}. This phenomenon will be more serious in the high frequency band \cite{alozie2022review}. On the other hand, the influence of long-distance transmission and space environment will also lead to the problems of high delay and large jitter in the communication process, by which the contradiction between timely adjustment and long delay of feedback adaptation mechanism, makes conventional error control difficult to work.

\subsection{Interconnection of Heterogeneous Networks}
SAGIN is composed of a variety of heterogeneous networks, and the environment and characteristics of each network are of greatly difference. For satellite space network, the satellite signals and transmission characteristics of different orbit positions are not the same. And for the terrestrial network, the different communication environment conditions are also various. In order to realize the integration, compatibility and convergence of SAGIN, it is necessary to have unity and similarity in the design of waveform regimes and communication standards \cite{alhashimi2023survey}. However, the above-mentioned interconnection characteristics of heterogeneous networks will inevitably make this goal more difficult.

\subsection{Load Resource Constraints}
Satellite system is with limited power resources. On the one hand, due to the rocket launch capability, the weight and size of the satellite are limited, which is constrained by the size of the solar panels. On the other hand, the power consumption of the payload is greatly limited by the thermal radiation capacity of the satellite \cite{wang2022conditional}. The larger the area of the radiator is, the stronger the heat dissipation capacity of the whole satellite is, and the higher the power consumption can be supported. However, due to the limitation of the profile size of the satellite and the influence of the satellite antenna, the size of the thermal radiator cannot be increased at liberty. With the development of satellite communication demand towards higher peak rate and more connections, the contradiction between the limitation of on-board power resources and the increase of transmission power. The improvement of on-board processing capacity will be further aggravated.

\subsection{Wide Area Transport Security}
SAGIN realizes wide-area coverage of users by means of satellites and HAPs. However, the wireless channel of satellite communication has the characteristics of openness and broadcast, which makes the information transmission channel uncontrollable The wireless link is more vulnerable to threats such as human interference, attack, eavesdropping and replay. Therefore, the development of SAGIN needs to solve the transmission security challenges under the circumstances of wide area coverage.

\section{Solution}
\label{Solution}
In order to meet the future application requirements of SAGIN in all fields and scenarios, and solve the difficulties and challenges faced by the current development of SAGIN, by studying the literature in recent years, the optimization strategies can be explained from the following aspects: architecture, performance, and AI.

\subsection{Solution with Architecture}
The update and iteration of the architectures is an important part of the communication technology development. It is also a popular research direction in the field of 6G communication. The following is mainly to introduce the ways to optimize architecture from two aspects: access mode and multi-level deployment.

\subsubsection{Access Method}
With the continuous development of satellite technology, the number of satellites in different orbits and the payload capacity have been significantly improved. The future satellite communication network will develop along the following three-stage evolution route:

Phase 1: The satellite uses bent pipe as the main working mode to support the ground base stations to access 6G CN through satellite return transmission. 

Phase 2: Although satellite bent pipe technology is used, various types of UE such as vehicles, ships and UAVs can be connected to the satellite to access the ground base station and then access the 6G CN. 

Phase 3: LEO satellites have spaceborne base station capabilities, and in the advanced phase, GEO can carry lightweight 6G core network elements (6G CN Lite) \cite{li2021cognitive}.

From the perspective of network access, in the first stage, UE accesses the ground base station and the satellite acts as a link back transmission. In the second stage, the UE accesses the satellite with transparent forwarding to connect the ground base station and the CN. In the third stage, UE accesses the CN through the spaceborne base station. With the update and iteration of access technology in the above three evolution stages, the technologies adopted will coexist for a long time, which can also be regarded as different access methods.

\subsubsection{Multilevel Deployment}
Cloud-based 6G satellite communication network deployment is more convenient and flexible.[] By constructing a 6G-oriented distributed cloud architecture, the user surface sinks to the edge node, which effectively extends the computing power of the cloud from the center to the edge, realizes rapid business processing and nearest forwarding, and meets a variety of application scenarios.

Therefore, the space-based and ground-based CN of 6G satellite communication network should also adopt multi-level deployment mode \cite{huang2022generalized}. For the space-based core network, the lightweight CN was deployed by GEO \cite{li2021cognitive}, while the network edge functions such as Mobile Edge Computing (MEC) and content distribution were mainly deployed instead of GEO constellation \cite{al2021survey}. Accordingly, the foundation CN is also deployed in multi-level data centers, that is, the center, region and edge three-level data center with base station computer room is as the infrastructure. Network elements can be deployed in the corresponding locations of the network according to the scene requirements.

\subsection{Solution with Better Performance}
There is an urgent need for low latency and high flexibility in many scenarios of today's SAGIN networks.

\subsubsection{Reduce Latency}
There is considerable room to improve the delayed performance of the existing SAGIN. The main reason for the high delay is that the CN is deployed on the ground, leading to the large round-trip delay of the user-satellite-ground CN \cite{xing2023earth}. One of the solutions is to improve the data offloading mechanism, and migrate the necessary core functions of the user network from the ground to the air or space, so as to reduce the round-trip delay of the user-satellite network.  

Develop the satellite base station capacity, strengthen the communication capacity between satellites, further balance the load of SAGIN network, reduce the time delay, and ensure the elastic contraction of system capacity. UAVs and other aircraft can expand the coverage area, optimize routing, and realize data diversion from ground base stations.

Updating existing centralized cellular networks to be distributed, i.e. some elements of the cellular network can be deployed closer to the RAN rather than centrally in the data center\cite{kazemifard2021minimum}, resulting in a local cellular system, is also a way to reduce latency.

\subsubsection{Increased Flexibility}
Hardware customization is the reason that restricts the flexibility of the SAGIN network. In the existing architecture, all the functions of RAN and some functions of CN are customized by hardware \cite{masur2022artificial}. Therefore, many functions cannot be designed flexibly according to the requirements and characteristics of platforms and application scenarios, which will also bring the negative impact of high deployment costs and high update and maintenance costs.  

In contrast, the trend of SAGIN networks is service-oriented, customizable, and lightweight. RAN no longer uses proprietary equipment, but uses service-based technology. Each function can be decomposed and refined. Cstomized specific functions can be combined according to different application scenarios and characteristics, providing a feasible solution for distributed deployment in SAGIN to form a distributed system.

\subsection{Integrated with AI}
In view of the problems existing in SAGIN mentioned above, AI can be integrated with SAGIN network in channel modeling, spectrum allocation and network structure optimization.

\subsubsection{Channel Modeling}
SAGIN is more complex than existing networks, with heterogeneity, self-organization and time-varying characteristics. Therefore, the idea of design and optimization of SAGIN is quite different from that of traditional relatively discrete networks. More technological innovations are needed to break through the development barriers. AI can be used to optimize the SAGIN network. For the channel modeling link, the data analysis ability of AI can be used to explore the feasibility and other characteristics of the new model \cite{huang2022artificial}. In the future, AI will play a more active role in SAGIN. In addition, AI-based physical layer techniques in SAGIN communication systems, such as deep learning-based massive MIMO\cite{chen2022survey}, etc., also contribute to improving network performance.

\subsubsection{Spectrum Allocation}
Spectrum has become a scarce resource, which is the inevitable result of the rapid increase of mobile devices. It is the focus of current attention to make full use of limited resources to provide high-speed and reliable services for the fast growth of mobile users. AI and big data have the ability to be applied to wireless channel resource allocation to better improve the way of spectrum allocation. Through ant colony algorithm, Bayesian reinforcement learning, deep learning and other means, the dynamic spectrum allocation strategy\cite{cullen2023predicting} was optimized, while the frequent channel switching was reduced, and the effectiveness of spectrum sharing was enhanced.

\subsubsection{Network Structure Optimization}
The deep integration of 6G with cloud computing, big data and AI has reached a consensus in the industry. AI will become the brain of mobile communication network. The integration of AI and SAGIN has great potential in the direction of resource utilization \cite{wang2020sfc}. It makes full use of the ground data center and air resources on-demand integration as a unified resource pool, and takes the cloud computing architecture as the underlying framework to build a more flexible, intelligent, efficient and open network system. It is conducive to improving the security of communication systems and better providing users with confidential data transmission while efficient and fast network design, analysis, optimization and management. Deep integration with AI for network structure optimization can become the drive for network intelligence and service intelligence.

\section{Conclusion}
\label{Conclusion}
In this paper, we summarize the research related to SAGIN that has attracted a great deal of attention for the past few years. Firstly, the basic characteristics of SAGIN network are summarized from the aspects of previous networks structure and physical layer characteristics. Then, the application scenarios, drive  and prospects of SAGIN network architecture are discussed in detail. Based on the summary of the existing work, this paper points  out the main technical challenges faced by the design of such converged networks, summarizses  the existing optimization strategies and prospects the future research directions worthy of further explorations.

\bibliographystyle{IEEEtran}
\bibliography{SAGIN.bib}

\end{document}